\documentclass[acmsmall]{acmart}

\AtBeginDocument{%
  \providecommand\BibTeX{{%
    \normalfont B\kern-0.5em{\scshape i\kern-0.25em b}\kern-0.8em\TeX}}}

\copyrightyear{2020} 
\acmYear{2020} 
\setcopyright{acmlicensed}\acmConference[MobileHCI '20]{22nd International Conference on Human-Computer Interaction with Mobile Devices and Services}{October 5--8, 2020}{Oldenburg, Germany}
\acmBooktitle{22nd International Conference on Human-Computer Interaction with Mobile Devices and Services (MobileHCI '20), October 5--8, 2020, Oldenburg, Germany}
\acmPrice{15.00}
\acmDOI{10.1145/3379503.3403555}
\acmISBN{978-1-4503-7516-0/20/10}

\usepackage{array}
\usepackage{dirtytalk}
\usepackage[utf8]{inputenc}

\begin{document}

\title{Medical Selfies: Emotional Impacts and Practical Challenges}

\author{Daniel Diethei}
\email{diethei@uni-bremen.de}
\affiliation{%
  \institution{University of Bremen}
  \city{Bremen, Germany}
}

\author{Ashley Colley}
\email{ashleycolley@gmail.com}
\affiliation{%
  \institution{University of Lapland}
  \city{Rovaniemi, Finland}
}

\author{Matilda Kalving}
\email{Matilda.Kalving@ulapland.fi}
\affiliation{%
  \institution{University of Lapland}
  \city{Rovaniemi, Finland}
}

\author{Tarja Salmela}
\email{tarja.salmela@ulapland.fi}
\affiliation{%
  \institution{University of Lapland}
  \city{Rovaniemi, Finland}
}

\author{Jonna Häkkilä}
\email{jonna.hakkila@gmail.com}
\affiliation{%
  \institution{University of Lapland}
  \city{Rovaniemi, Finland}
}

\author{Johannes Schöning}
\email{schoening@uni-bremen.de}
\affiliation{%
  \institution{University of Bremen}
  \city{Bremen, Germany}
}

\renewcommand{\shortauthors}{Diethei et al.}

\begin{abstract}
Medical images taken with mobile phones by patients, i.e. medical selfies, allow screening, monitoring and diagnosis of skin lesions. While mobile teledermatology can provide good diagnostic accuracy for skin tumours, there is little research about emotional and physical aspects when taking medical selfies of body parts. We conducted a survey with 100 participants and a qualitative study with twelve participants, in which they took images of eight body parts including intimate areas. Participants had difficulties taking medical selfies of their shoulder blades and buttocks. For the genitals, they prefer to visit a doctor rather than sending images. Taking the images triggered privacy concerns, memories of past experiences with body parts and raised awareness of the bodily medical state. We present recommendations for the design of mobile apps to address the usability and emotional impacts of taking medical selfies.
\end{abstract}



\begin{teaserfigure}
  \includegraphics[height=3.7cm]{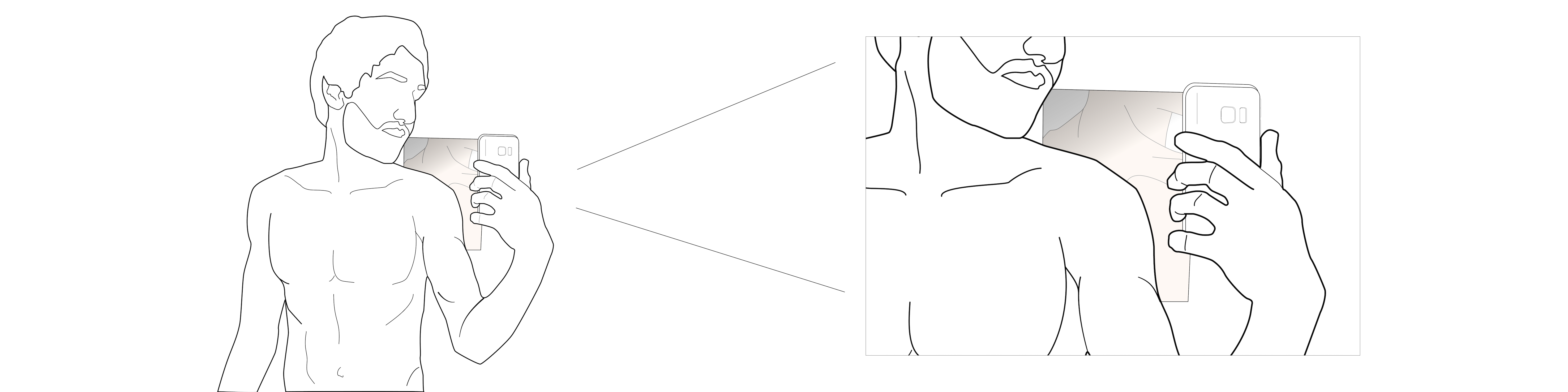}
  \caption{Taking a selfie to monitor skin lesions on the back, using a mirror.}
  \Description{A person taking a selfie to monitor skin lesions on the back, using a mirror.}
  \label{fig:teaser}
\end{teaserfigure}

\maketitle

\section{Introduction}

The incidence of skin cancer has reached epidemic proportions in white populations and the trend is still rising~\cite{kroemer2011mobile}. Currently, between 2 and 3 million non-melanoma skin cancers and 132,000 melanoma skin cancers occur globally each year. One in every three cancers diagnosed is a skin cancer and, according to the Skin Cancer Foundation, one in every five Americans will develop skin cancer in their lifetime~\cite{WHOskincancer}. Early detection and treatment are essential in reducing mortality. While the technical equipment commonly used in this context has previously comprised expensive stereomicroscopes and digital dermoscopy systems, teledermatology has shown to be more cost effective~\cite{trettel2018telemedicine, livingstone2015assessment, lee2018recent} and at least as accurate in diagnosis~\cite{markun2017mobile, wang2017diagnosis} compared to face-to-face consultations. Mobile teledermatology has good diagnostic accuracy for skin tumours~\cite{kroemer2011mobile,lamel2012application,kirtava2016health}. Due to the low-cost infrastructure it is a convenient tool, especially in resource-limited settings. Examples of the succesful implementation of teledermatology services are reported from the Netherlands with 130531 consults between 2006 and 2015~\cite{tensen2016two} and from the Brazilian state of Catarina with 83100 consults from 2014 to 2018~\cite{von2019creating}.

\begin{figure}[b]
  \centering
  \includegraphics[width=0.9\columnwidth]{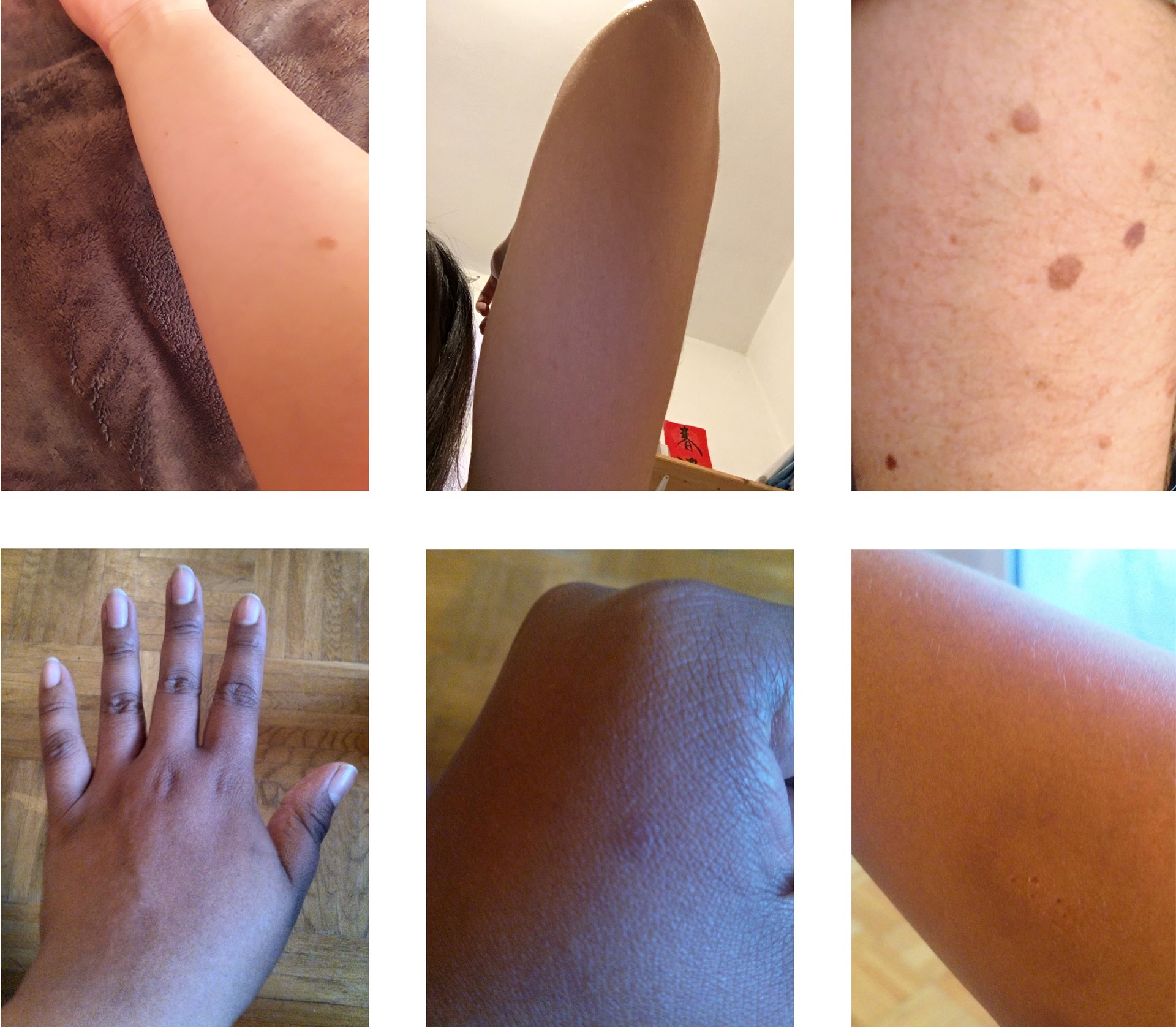}
  \caption{Example images of skin photos of study participants. Note: In the study protocol, the researchers did not have any access to participants' photos. The above photos were specifically volunteered by participants for publication.}
    \Description{Photos of participants' arms and hands, some of them with moles.}
  \label{fig:example_photos}
\end{figure}

While the technology has evolved rapidly~\cite{lee2018recent}, the emotional impacts of sharing medical photos between patients and doctors have not been considered in literature to the same extent. 

Sending sensitive medical images through an abstract communication channel that is not as tangible as face-to-face communication requires patients to build a high level of trust in the technology. This has implications for the design of applications and systems through which patients collect and distribute sensitive medical photos, and potentially receive back life-changing diagnosis. In this context, we address the concept of affective atmospheres ~\cite{lupton2017affective}, whereby the combination of technology, spaces and actors contribute to the overall affective experience of an individual using a system. When dealing with health and technology use, people can be vulnerable and are exposed to strong feelings. 



When there is no physical presence of a doctor, sensory engagement with the patient, an essential part of developing trust, is lacking and has to be compensated for. Interactions between humans and digital devices can be personal and intimate experiences ~\cite{fullwood2017my,hemmert2011intimate,venta2008my}. In the medical context, the visual, aural and haptic aspects of digital devices become increasingly important to support the emotional impacts of their use~\cite{lupton2017affective}. Recent work exploring the use of digital resources for HIV, has identified that such considerations for providing emotional support are currently missing from online medical services~\cite{Singh:2019}. We propose that in situations where digital imaging for the purposes of disease detection takes place, a particular atmosphere is created that is dependent on, for example, the body parts being digitally documented by an individual for disease detection purposes; the current level of knowledge of the medical state of the particular body part being photographed for diagnosis; the existing relation with one's body; the presence and form of a camera used to photograph one's body; the presence or absence of other humans in the space; the familiarity or foreignness of the space and; the ways in which the images are to be forwarded to medical experts. Moreover, we propose a possibility that the usage of cameras of personal smartphones to take medical photographs of one's own (private) body parts can create negative emotional states such as insecurity and discomfiture. This has a strong connection with the digital documenting of specific body parts that are considered as more sensitive, vulnerable or private to be taken pictures of.




The aim of our work is to explore the area of medical selfies, i.e. images taken of one's self to record the state of a medical condition, in regards to their emotional impacts and practical challenges. We address the following research questions 
\begin{itemize}
    \item \textbf{RQ 1}: What are the emotional impacts associated with the process of taking medical selfies and how do they differ between body parts?
    \item \textbf{RQ 2}: What are the physical and practical challenges people face when taking medical selfies and how can mHealth apps be designed to overcome them?
\end{itemize}

The joint observation of the emotional together with the physical dimensions enables us to derive design guidelines specific to the intimate nature of medical selfies.

In this paper, we report our findings from an online survey (n=100) and an in-depth user study (n=12) on how people experience taking medical images of their body. Survey respondents expressed that they mostly prefer to not send medical photos of intimate body parts but instead rather see a doctor. However, for other body parts and less severe skin conditions, such as rashes, respondents stated that selfie images were the preferred analysis mode. Participants in the user study found it physically challenging to take photos of their buttocks and shoulder blades and expressed most emotional discomfort with genital and buttock photos. Inspecting their bodies as part of the study procedure raised their awareness of skin features, e.g. dryness and moles. Self-examination made participants check on body parts they would not normally look at in their daily life. Participants were concerned about their appearance in medical selfies, even though they were aware of the sole medical purpose. We provide design guidelines (Table \ref{tab:guidelines}) that address practical challenges, e.g. correct device positioning and image acquisition, and emotional dimensions such as aesthetics and data privacy.
\section{Related Work}
%

As relevant prior work, we firstly summarise the current state of knowledge on selfies in general, focusing on emotional and practical aspects. We then briefly reflect on the trend towards online medical care, so called telemedicine, as one of the main drivers for medical selfies. We then detail related work looking at the medical selfie from a variety of angles. Finally, we highlight the contribution of our work to the current state of knowledge.

\subsection{The Selfie Phenomenon}

Nowadays, taking selfies, i.e. photographs of oneself, and sending them to friends, or posting them more publicly, is a common practice across much of society~\cite{bohmer2011falling}. 
There is a large body of prior work on selfies, exploring a broad variety of the motivations and effects of the phenomenon, e.g.~\cite{senft2015selfie, barry2017let, lottridge2017selfie}.
Sung et al.~\cite{sung2016we} identified four motivations for posting selfies on social media,  attention seeking, communication, archiving, and entertainment. Higher levels of selfie activity have been reported as an indicator of higher body satisfaction and narcissism~\cite{cohen2018selfie, sung2016we}. However, high investment in selfie images correlates with decreased body satisfaction~\cite{lonergan2019me,cohen2018selfie}. Higher investment in the images can take the form of elaborate staging (e.g. environment and make up), time spent selecting photos for publication or effort in photo manipulation~\cite{mclean2015photoshopping}. Cohen et al. note that over half of the participants in their study reported manipulating their selfies through image editing "sometimes" to "very often"~\cite{cohen2018selfie}.

Surprisingly, we were unable to find prior works specifically exploring the ergonomics of selfie taking. However, Arif et al.~\cite{Arif2017} have reported on the usage of selfie-sticks, noting their general unergonomic design and recommending improvements such as a wider diameter grip. Medical papers have reported occurrences of `selfie elbow', caused by the excesses of holding the smartphone at arms length~\cite{chugh2016selfie}. More seriously, distraction during selfie taking has caused injury or even death~\cite{wikipedia2019selfiedeath}. Selfie images may not present a fully accurate representation, being affected by lens distortions~\cite{ward2018nasal} and mirroring effects~\cite{bruno2015selfie}. Issues of mirroring effects have been explored by Bruno et al.~\cite{bruno2015selfie}, who report a left cheek bias for 'standard' selfies. 



\subsection{Online Self-Diagnosis and Telemedicine}

As an alternative to the traditional visit to a medical clinic, patients may seek self diagnosis and care using freely available online information sources i.e. `Googling symptoms'~\cite{lupton2018digital}. Similarly many online discussion forums exist where patients may seek community diagnosis. For example Reddit hosts a number of specific ailment related forums including r/DermatologyQuestions\footnote{https://www.reddit.com/r/DermatologyQuestions/}, where photos and descriptions of skin-related issues are posted, discussed and in many cases diagnoses by others users are made. Brady et al.~\cite{brady2016you} highlight that such forums enable the building of trust in particular advisors, through observation of their postings over time. Others have reported that the use of internet self diagnosis can reduce satisfaction with medical professionals, when they are later consulted~\cite{robertson2014my,sjostrom2019primary,lupton2015s}. Patients' inability to manage internet sourced information and its potential inaccuracy have been highlighted as the main reasons for patient-clinician conflict~\cite{sjostrom2019primary}.

Professional clinician use of telemedicine has seen rapid growth in recent years, achieving a US market size of \$26.5 billion in 2018, growing to a predicted \$130 billion by 2025~\cite{hcweekly2019guide}. The domain of telemedicine is rather broad, including e.g., mobile health platforms, real-time interactive services, store and forward services, and remote patient monitoring and addressing e.g., dermatology, oncology, psychiatric conditions and post surgery follow up. However, technology adoption challenges amongst medical staff and patients, particularly elderly patients, present barriers to the success of telemedicine services.~\cite{scott2018evaluating}. Suggested approaches to address the identified issues include staff training and alternating telemedicine and personal patient-to-provider interactions~\cite{scott2018evaluating}. Teledermatology, a subset of telemedicine, is the practice of dermatology using information and communication technologies. Recently, Abbott et al.\cite{abbott2020telederm} summarised 16 teledermatology reviews and concluded that teledermatology is comparable to traditional in‐person methods of delivering dermatological care. Most reviews state that teledermatology is similar in terms of diagnostic accuracy \cite{lee2018recent, bashshur2015empirical, chuchu2018teledermatology} and management and patient outcomes \cite{bashshur2015empirical, chuchu2018teledermatology, lee2018teledermatology} with the exception of two studies that report lower diagnostic accuracy for teledermatology \cite{finnane2017teledermatology, warshaw2011teledermatology}. Focusing on smartphone apps to address melanoma detection, Kassianos et al.~\cite{kassianos2015smartphone} identified 39 such apps that included features such as education (e.g. in taking images of lesions), classification, risk assessment and monitoring change. Whilst some of the apps provided the possibility to send images to medical professionals, generally the apps were lacking clinical validation. Similar findings are echoed by Lupton and Jutel~\cite{lupton2015s}, who also note the routine use of disclaimers in such apps undermines trust in their diagnosis.

\subsection{Medical Selfies}
Prior work on medical selfies has addressed both their explicit function as a medical tool, particularly in the area of dermatology, as well as other indirect motivations and benefits for their use.

A concise overview of the medical selfie is provided by Burns~\cite{burns2015digital}, highlighting their use to document ailments with visual pathology in areas such as dermatology, rheumatology, ophthalmology and burns treatment~\cite{burns2015digital}. In a meta review of skin self-examination (SSE) Yagerman et al.~\cite{yagerman2013} highlight its benefits in reducing melanoma incidence and mortality. Patient demographics are noted as influencing willingness and ability to perform SSE. Older individuals may have limited visual acuity and poor flexibility, which, together with the background skin lesions that develop with age, result in low confidence in performing SSE\cite{yagerman2013}. For afflictions with transient symptoms taking selfies can provide proof or a memory to aid diagnosis and discussion with clinicians~\cite{bbtimes2019selfie}.

Several works have discussed legal and privacy implications of using medical selfies~\cite{burns2015digital, ray2015, zampetti2016medicine, chao2017smartphone}. Problems arise when medical practitioners provide guidance on taking selfies, when they receive images in emails or WhatsApp messages and when patient-taken images are included as part of medical records~\cite{burns2015digital, zampetti2016medicine}. Further concerns of data ownership and privacy are reported by Chao et al.~\cite{chao2017smartphone}. On the other hand, Ray et al.~\cite{ray2015}, note that images taken by patients are not subject to the strict data protection rules of medical practitioner taken photographs. 

An overview of the use of smartphones to take dermoscopic images is presented by Ashique et al.~\cite{ashique2015clinical}, who highlight consistent lighting as the main challenge in reliable assessment. Prior work has reported positively on patients' ability to take high-quality dermoscopic images at home, using smartphones~\cite{manahan2015pilot}. Taking selfie images of the back is a notable problem area, with solutions using two mirrors or a selfie-stick being proposed~\cite{criscito2016selfie}. Whilst processes including the post-analysis of such images by medical practitioners have been shown to be beneficial, automated app-based evaluation have raised safety concerns~\cite{rat2018use}. In a recent work, Ngoo et al.~\cite{ngoo2018fighting} provide an overview of melanoma related apps in the app stores, reporting that the most common target of apps is to support users in monitoring their moles over time. Such apps may include self, automated and medical practitioner assessment of taken images, e.g.  Miiskin\footnote{https://miiskin.com/app/} Molescope\footnote{https://www.molescope.com/} and Skinvision\footnote{https://www.skinvision.com/}. Common app user interface features include reminders to re-asses lesions and side-by-side presentation of historical and current images of the same lesion. A key challenge in lesion imaging is ensuring correct and repeatable camera positioning, this has so far been addressed though automatic shutter activation, which may be enhanced with audible positioning guidance~\cite{Diethei:2018}.



In addition to the direct role of medical selfies in supporting a patients physical treatment, several works have investigated positive impacts of the process on patients' mental state~\cite{tembeck2016, yagerman2013, ray2015, burns2019}.
Tembeck~\cite{tembeck2016} discusses the use of medical selfies as a form of self expression, enabling individuals to publicly identify themselves as living with illness, and to highlight its centrality to their daily life. A common finding from several studies is the positive effect of selfie taking in encouraging patients to take ownership of their condition and treatment~\cite{yagerman2013, ray2015, burns2019} 

\subsection{Contributions}
From the related work, we learn that selfie based solutions will play an increasingly important role in the prevention, diagnosis and treatment of skin lesions. Technical issues, e.g. image quality, have have been well addressed, and legal and privacy issues at least identified as requiring further study. Several threads related to emotional and experiental aspects of taking medical selfies have been opened by prior work, and remain largely unresearched. For example, there is little research on the presence of emotional impacts similar to those of normal selfies, ownership of one's condition and treatment, and trust in those assessing the images, be they medical professionals, unknown individuals in an online forum or algorithms in a mobile app. As a contribution, we aim to take steps towards understanding the interconnections between these underlying experiences, and deliver knowledge enabling the creation of improved patient experiences.

\section{Method}
We collected two complementary datasets. Firstly, we collected data from an online survey (n=100), aiming to gain a broad overview of the issues surrounding the topic. Secondly, to gain deeper insights, we conducted an in-depth user study (n=12). 

\subsection{Online Survey}
We developed an online survey exploring experiences and attitudes towards teledermatology, particularly aspects requiring taking photographs of one's own body. As well as demographic data and information on the participants' general approach to technology usage, data on feelings of body esteem were collected e.g., `I feel good about my body'. Gender specification was voluntary and followed the guidelines by Spiel et al.~\cite{spiel2019better}. All participants reported their gender.
The survey included 5-point rating scale questions, e.g. to address comfort levels of taking photos for different body parts, as well as two open questions, (1) in which situations respondents prefer to have a face-to-face examination with a physician and (2) in which situations respondents prefer to send photos to physicians.


The survey was distributed via the Amazon Mechanical Turk crowdsourcing marketplace, with the only participation criteria being that the respondents' location was in the US. Participant compensation of 1\$ was provided based on the estimated task time of 7 minutes and the average minimum wage \cite{minimumwageg2019}.
The survey resulted in data from 100 participants (\textit{M$_{Age}$} = 34.2, \textit{SD} = 9.1; 43\% female, 56\% male, 1\% non binary). The average completion time was 4.7 minutes (\textit{SD} = 3.6; range: 1.6 to 31.0 minutes). We received 111 responses, of which we excluded eleven due to missing data. The average word count for each of the two open questions in the survey was 11.23 words (\textit{SD} = 15.27). The survey data is available in the supplementary material.

Respondents' textual responses to open questions were analyzed using an open coding approach ~\cite{strauss1987}. One researcher defined the codebook, two coders independently evaluated each response, and  a third researcher arbitrated disagreements between the coders. Answers were coded such that each answer could produce codes in multiple categories. 

\subsection{In-Depth User Study}
For deeper qualitative insights we conducted an interview based study with twelve participants. Participants were recruited through local online classified ads and flyers placed in university cafeterias. The intimate nature of the photos led to a challenging recruitment process. Due to the rather complex study procedure, two participants dropped out after the briefing session. The recruited participant sample was gender-balanced (6 male, 6 female) with half residing in Finland and half in Germany. The mean participant age was 29.66 years (\textit{SD} = 8.63, range 20-41) and they were all familiar with using smartphones (\textit{M$_{experience}$} = 7.42 years, \textit{SD} = 2.31). Ten participants were university students, one was in a full-time working position and one was on parental leave. Participants were rewarded with a cinema ticket as compensation for their participation.

The in-depth user study process consisted of the following stages:
1) Initial briefing and equipment setup, 2) Taking medical selfies, video diary and questionnaire, 3) Follow up interview. A key tenet of the study procedure was maintaining the participants' privacy. Thus, taken images were at no point shown to the researchers.

At an initial meeting with the participants, the purpose of the study was explained and participants were instructed on the procedure. Informed consent, developed under guidance from the university's  ethics advisor, was signed by the participant. When required, participants were assisted in the installation a voice recording app, required for part of the study. To ensure the privacy of taken images was maintained, participants were given instructions on how to disable the automatic upload of photographs to cloud services, such as Google Photos and Apple iCloud. 

Later, in their own homes, at a time and place they felt comfortable with, participants followed a set of printed instructions to capture photos of their body parts. For the task setting, participants were asked to imagine that they had been requested to take photographs by a family doctor, showing moles or other skin conditions. A list of body parts to be photographed was provided. On each body part participants were instructed to capture skin features, such as scars, moles and veins, irritations. If no salient features were present, participants were asked to capture the full body part. During the capture process, individuals were instructed to think aloud and the audio was recorded on their device through a voice recording app running in the background. While the body parts needed to be fully visible, we made it clear that it was not necessary to completely undress. If a participant felt uncomfortable taking photos of any of the body parts, they were free to skip them. We asked participants to give an explanation for their reasoning in this case. Participants were asked to retain the images taken until after the final interview. Figure \ref{fig:example_photos} shows a collection of example photos which were volunteered by participants for publication after the completion of the study.


Immediately after participants had taken photos of all the body parts on the list, we asked them to record a selfie video in a diary manner to give feedback on the experience, guided by a set of questions (Table \ref{tab:selfievideoquestions}). Finally, participants completed an online questionnaire that covered demographics, technological affinity~\cite{karrer2009technikaffinitat}, how physically demanding it was to take the photos. 

\renewcommand{\arraystretch}{1.5}
\begin{table}[t]
\centering
\begin{tabular}{m{\columnwidth}}
\hline
How did it feel to take photos of your body for medical purposes?  \\  
\hline
While taking the photos, have you recognized and photographed features on
your skin you were already concerned about before the study? If yes, please elaborate which features there were and how you felt
taking photos of them. \\
\hline
While taking the photos, did you discover new skin features you are now concerned about? Was there anything that surprised you? If yes, how did it feel discovering and photographing these features? \\
\hline
Describe the experience (taking photos of your body for medical purposes) in terms of privacy and intimacy. \\
\hline
Have the photos changed the perception on your bodily medical state? If yes, how? \\
\hline
\end{tabular}
\caption{Questions study participants answered in a selfie video immediately after they took the photos. }~\label{tab:selfievideoquestions}
\end{table}

The video and audio recordings were sent via email to the researchers. The photos taken were not sent or shown to the researchers.
The following day, a semi-structured interview was conducted, either at the participant's home or at the university. In addition to preset questions (Table \ref{tab:interviewquestions}), individual questions were included, based on pre-analysis of the individuals' responses to the questionnaire. The interview was audio-recorded and transcribed.

\begin{table}[t]
\centering
\begin{tabular}{m{\columnwidth}}
Do you feel comfortable sharing the photos with a doctor you know? How about a doctor you do not know?  \\  
\hline
Could you imagine having an app to diagnose diseases on the photos you captured? \\
\hline
How do you compare the atmosphere of intimacy between taking the photos at home and being at a dermatologist screening? \\
\hline
If you have used a mirror: Was it useful? \\
\hline
Do you have any experience in sharing medical photos with friends or family? \\
\hline
How does it feel having the photos on your device? \\
\end{tabular}
\caption{Common interview questions asked one day after the capture session. Additional questions were asked individually based on previous statements.}~\label{tab:interviewquestions}
\end{table}

As with the survey analysis, we used an open coding approach with three researchers for the audio, video and interview transcriptions. Since both Finnish and German researchers were involved in the process, the analysis was done based on the English transcriptions. The 36 transcriptions consisted of approximately 18 000 words. We provide the transcripts as supplementary material.

\subsection{Data Privacy \& Ethics}
Since the nature of our study required participants to take highly sensitive photos, we took a lot of care in minimizing discomfort and ensuring data privacy during the study. First and foremost, we did not see any of the photos the participants took at their home and communicated this clearly at the first briefing. Furthermore, we explicitly advised to turn off cloud services to prevent the photos being uploaded to the internet. While an image quality rating would have been helpful in assessing the clinical relevance of the images, we set the focus of our study on emotional and practical dimensions. Therefore, we decided to be as least invasive as possible to make sure participants acted naturally. Moreover, for ethical reasons, we advised participants that they could skip body parts should they feel uncomfortable capturing them. Only a minority of the participants actually skipped photos, which shows that there were few concerns with the legitimacy of the study.

\section{Findings}


In this section, we report the combined findings from our online survey and in-depth user studies. We refer to individuals as survey \textit{respondents} and user study \textit{participants}, e.g. P2 as the second participant of the in-depth study and R36 as respondent 36 of the survey. The most common themes of the in-depth study are listed in Table \ref{tab:themes}.
Specifically we report on prior experiences with sending medical selfies, preferences for in-person consultations, ergonomics of taking the images, emotional effects, impacts of the taking selfies and privacy concerns. 

\begin{table}[b]
\centering
\begin{tabular}{ m{6cm} | l }
\textbf{Theme} & \textbf{\# of mentions} \\
\hline
Data Privacy & 40\\
Discovering skin lesions & 30\\
Memories of past events associated with body parts & 29\\
Skin self examination & 17\\
Relationship to doctors & 16\\
Assistance used (e.g. mirror)  & 15\\
Prefer to see a doctor in person & 13\\
Aesthetics (both positive and negative) & 13\\
\end{tabular}
\caption{In-depth user study results: Most prominent themes identified from audio, video and interview transcriptions.}~\label{tab:themes}
\end{table}

\subsection{Prior Experience with Sending Medical Selfies}
Half of the study participants reported having previously sent medical selfies. One participant (P2) mentioned sending a photo of a lesion to a friend, who advised her to see a doctor. The lesion was then diagnosed as a benign melanoma. Another participant described her regular exchange of medical selfies with her friends and family: \say{[W]e are always analyzing with a friend or one of my sisters, like, hmm, what is it this time?}, (P3).

\begin{figure*}[t]
\includegraphics[width=1\columnwidth]{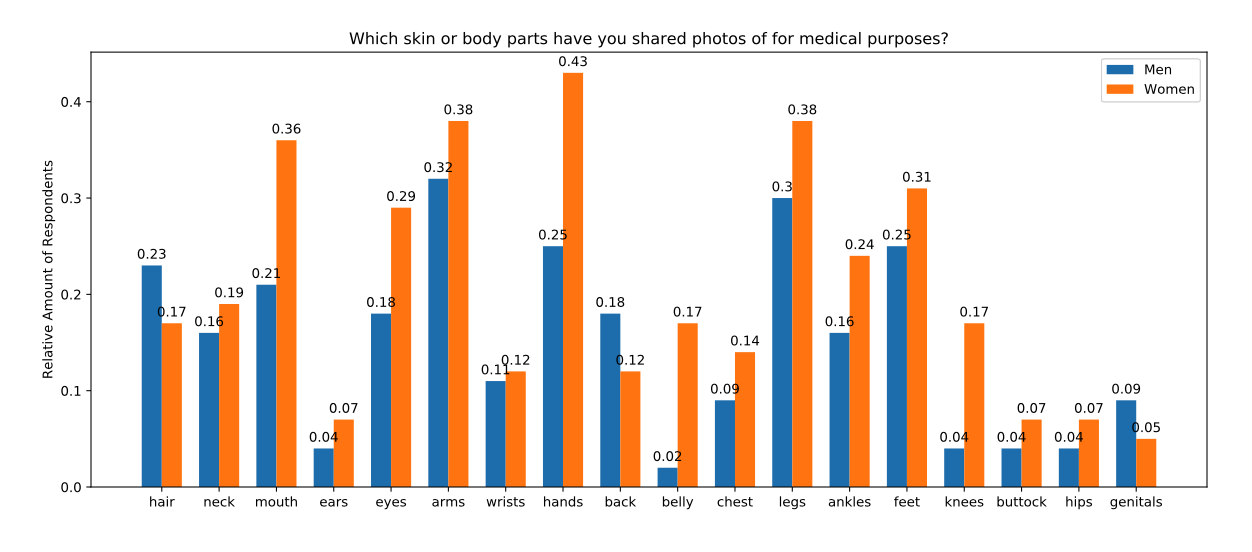}
\caption{Survey results: Proportion of male (n=57) and female (n=42) respondents that have shared photos for medical purposes by body part.}~\label{fig:exp_shared_bps_gender}
\Description{Bar chart with the proportion of respondents that share photos of the following body parts: hair, neck, mouth, ears, eyes, arms, wrists, hands, back, belly, chest, legs, ankles, feet, knees, buttock, hips, genitals.}
\end{figure*}

From the survey, 88\% of respondents reported having shared photographs of their body parts for medical purposes at least once. 9\% of respondents reported having done so five times or more. A $\chi^2$ test of independence revealed no significant effect of gender on the history of sharing medical selfies, $\chi^2$ (4, \textit{N} = 99) = .52, \textit{p} > .05. To explore differences in medical selfie content, we conducted an independent samples t-test on the effects of gender on the range of body parts respondents reported having shared (Figure \ref{fig:exp_shared_bps_gender}). In general, female respondents shared images of more body parts (\textit{M$_{body\_parts}$} = 3.71, \textit{SD} = 2.99) than male (\textit{M$_{body\_parts}$} = 2.65, \textit{SD} = 1.89) respondents (\textit{t}(97) = -2.169, \textit{p} < .05). Respondents (n=57) had the most experience in sharing photos of arms (35\%), legs (34\%) and hands (32\%). For particular body parts, no differences in the frequency of sharing was found between male and female participants (all Bonferroni-Holm adjusted \textit{p}-values > .18). Instant messaging was the dominant tool used for image sharing (72\%), followed by email (18\%), cloud services (3\%) and others (7\%).

Respondents reported sharing medical selfies with a broad set of those close to them, spouse/partner (54\%), family (62\%) and friends (43\%). Only 2\% of participants reported sharing such selfies with a doctor. Advice given by the confidants based on the medical selfies was evenly divided between recommendations to see a doctor (51\%) and not to see a doctor (43\%). The given advice was followed by 84\% of respondents. 


\subsection{In-Person Consultation vs. Sending Photos to a Doctor}
Almost two thirds (63\%) of the survey respondents said that conditions related to their skin were suitable for diagnosis by sending photos to their doctor. Here, saving time was a commonly mentioned justification, e.g. \say{sending a photo to someone is quicker. You won't waste time driving and doctor fees} (R34). Considering intimate body parts, 41\% of the respondents preferred that they would be checked by a doctor in person, whilst only a minority (5\%) stated a preference for taking photos and sending them for diagnosis. A typical comment expressing preference for an in-person consultation being, \say{genitals, breasts, anything that wouldn't usually be photographed or sent, photos I would be embarrassed leaking out into the internet} (R14). A similar preference ratio was reported in the study with only two participants (17\%) preferring to take photos of intimate body parts at home, rather than visiting a doctor.

For many respondents the type and location of the medical condition affected their preference for interaction mode. Respondents stated that they would prefer to send photos for less serious conditions (mentioned by 30\% of respondents), but would rather see a doctor for a serious condition (mentioned by 19\%).  Uncertainty of a condition's seriousness was seen as a justification to send photos (mentioned by 8\%), e.g. \say{for a dermatological issue that is most likely nothing, but I want to double check that it's nothing.}  (R7) and \say{maybe a mole that looked strange but I wasn't sure if it needed attention} (R36).

For the most commonly mentioned skin conditions (Table \ref{tab:skin_doc_app}), the preferred analysis mode was through selfie images. Respondents managed potential feelings of embarrassment in different ways. Some did not wish to take photos of sensitive conditions or body regions (mentioned by 4\%), e.g. \say{I would prefer to send photos if it was something not too embarrassing or uncomfortable to share.}(R44). On the other hand, others (5\%) wished to avoid showing and discussing such conditions and regions with a doctor in person.


\begin{table}[t]
\centering
\begin{tabular}{ l | l | l }
\  & \textbf{Prefer Doctor} & \textbf{Prefer App} \\
\hline
Rash & 11 (26\%) & 31 (74\%) \\
Cut & 1 (12.5\%) & 7 (87.5\%) \\
Bruise/Bruising & 0 (0\%) & 7 (100\%) \\
Eczema & 0 (0\%) & 4 (100\%) \\
\end{tabular}
\caption{Survey results: Preferences for diagnosis mode by skin condition.}~\label{tab:skin_doc_app}
\end{table}

When asked \say{I would trust a diagnosis based on medical photos from...}, a $\chi^2$ test of independence revealed that respondents have higher trust in a diagnosis by a doctor (69.5\% agree or strongly agree) compared to an AI-based diagnosis (23\% agree or strongly agree), $\chi^2$ (4, \textit{N} = 99) = 63.67, \textit{p} < .001. Moreover, a doctor they know (89\% agree or strongly agree) is preferred over a doctor they do not know (50\% agree or strongly agree), $\chi^2$ (4, \textit{N} = 99) = 47.81, \textit{p} < .001. Our user study revealed that a majority of participants (58\%) did not have any preference 
between sharing medical selfies with a doctor they know vs. a doctor they do not know. Three participants (25\%) preferred a doctor they knew, while two (17\%) said they would rather send images to an unfamiliar doctor. One participant made the distinction based on the intimate nature of the body part: \say{if it's not the private body parts, I feel it's fine to even share with a doctor I don't know. [...] I mean, at the first time we can meet in person, so I know his face and the personality of him and then further I share} (P7). Another participant expressing his opinion, \say{the unfamiliar may be the nicer one, yeah, in a funny way, I wouldn't know who's receiving, they just open it, and they won't know anything else about me} (P3). The same participant continuing, \say{it may be nicer If I knew the person, so I won't have the feeling that now I've sent these photos and now I have to meet him [afterwards]} (P3).

The attitude towards an AI-based diagnosis from an app on their mobile device differed among participants. While a majority (58\%) could imagine using an app to diagnose a condition from medical selfies, two participants added the provision that a doctor should decide in the end (P4, P5). Five participants (42\%) were critical towards app based automated diagnosis, preferring a traditional diagnosis from a doctor.

\subsection{Ergonomics and Image Quality}
A few survey respondents were conscious of potential misdiagnosis based on selfie images, commenting e.g., \say{[I would see a doctor for] anything that looked worse in a photo than in person, so things where I just couldn't get a clear and direct image or one that was clear and direct to the viewer...} (P47).

In general, participants used different strategies to reach the targeted body parts with the camera. For example, to capture the soles of the feet, P7 sat with their foot on a table, held in position with their non-camera hand, while P1 and P10 needed to sit down to take the photo. The study tasks revealed challenges in taking photos of some body parts, particularly the ones on the back of the body, i.e. buttocks and shoulder blades (Table \ref{tab:study_bodyparts}). Here, some participants employed a mirror, one participant describing standing with their back facing the mirror, placing their hand behind their back and pointing the smartphone's front camera towards the mirror (Figure \ref{fig:teaser}). Nine participants (75\%) reported they had difficulties capturing images of the shoulder blades,  e.g. stating, \say{that [shoulder blades] was definitely the hardest area to capture} (P1). On the other hand, a few participants explicitly mentioned the ease of taking photos of the shoulder blades, e.g. \say{right shoulder, um, this feels amusingly easy now at the end} (P3). Potential reasons for the differences are the use of a mirror and the individual's flexibility level. 

\begin{table}[t]
\centering
\begin{tabular}{ m{3.8cm} | m{2.8cm} | m{2.8cm} | m{2.8cm}}
\  & \textbf{Photo taken} & \textbf{Mirror used} & \textbf{Practical issues} \\
\hline
Hands & 12 (100\%) & 0 (0\%) & 2 (17\%)\\
Forearms / Underarms$^{*}$ & 12 (100\%) & 0 (0\%) & 4 (33\%) \\
Nostrils & 12 (100\%) & 4 (33\%) & 4 (33\%)\\
Belly & 12 (100\%) & 1 (8\%) & 1 (8\%) \\
Buttocks & 11 (92\%) & 1 (8\%) & 6 (50\%)\\ 
Genitals & 9 (75\%) & 0 (0\%) & 3 (25\%)\\
Soles of the feet & 11 (92\%) & 1 (8\%) & 2 (17\%) \\
Shoulder Blades & 11 (92\%) & 9 (75\%) & 9 (75\%) \\  
\end{tabular}
\caption{Study results: Body parts vs. photos taken, mirror used and practical problems (n=12). \newline $^{*}$ Due to a translation error, German participants captured their forearm, while Finnish participants captured their armpit.}~\label{tab:study_bodyparts}
\end{table}

Following our study protocol, the researchers did not see any of the images taken by the participants and thus were unable to make any assessment of the image quality, e.g. if the feature of interest was blurry. However, several participants described image quality issues when reviewing their own images. Three participants (25\%; P1, P7, P9) had problems with camera focus while capturing their belly, shoulder blades, hands, nostrils and buttocks. Two participants (17\%; P4, P8) struggled to find appropriate lighting conditions for taking the photographs, suggesting \say{it would be nice to provide more information while taking the photos, for example, if the light is good} (P7).

\subsection{Emotional Impacts of Medical Selfies} As well as practical challenges in taking the images of their own bodies, participants were also affected by emotional issues.
Participants in the study were willing to take photos of most body parts, the only exception being the genitals, which three participants (25\%) were not comfortable capturing (Table \ref{tab:study_bodyparts}). Interestingly, only one of the participants did not take the buttock photos. This may have been by accident, as no reasoning was given, and the the participant did photograph the genital area, which is usually the most sensitive body part. One participant took photos of their thighs instead of the soles of the feet and one participant skipped the last two photos of shoulder blades for unknown reasons. 

\begin{figure*}
\includegraphics[width=1\textwidth]{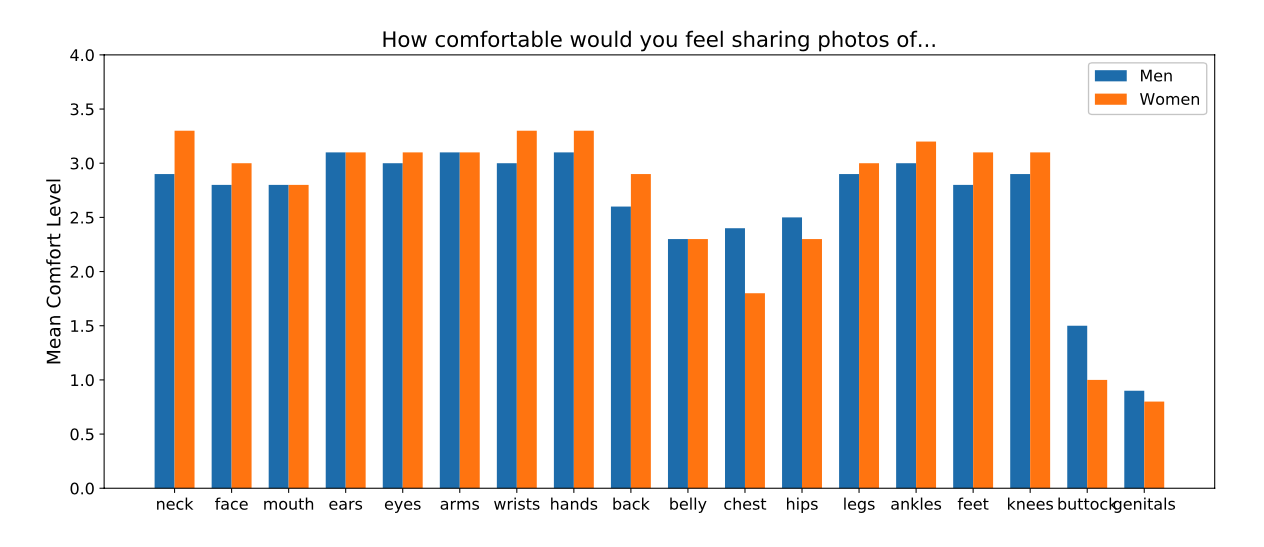}
\caption{Survey results: Mean comfort level of male (n=57) and female (n=42) respondents to share medical selfies by body part.}~\label{fig:comfort_bps_gender}
\Description{Bar chart with comfort levels of the following body parts: hair, neck, mouth, ears, eyes, arms, wrists, hands, back, belly, chest, legs, ankles, feet, knees, buttock, hips, genitals.}
\end{figure*}

Survey respondents were least comfortable with photographing the buttocks and genitals (Figure \ref{fig:comfort_bps_gender}). Additionally photographing the belly, chest and hips appeared to raise feelings of discomfort for some respondents. To test whether the independent variable gender (male/female) had an effect on the dependent variable photography comfort level (0-4), we conducted a MANOVA. Male and female survey respondents reported similar levels of comfort for individual body parts (\textit{F}(18,80) = 1.491, \textit{p} > .05, Wilk's 
$\lambda$ = .749). In general, buttocks (\textit{M}$_{comfort}$ = 1.31, \textit{SD} = 1.40) and genitals (\textit{M}$_{comfort}$ = .90, \textit{SD} = 1.18) were the most uncomfortable body parts to be shared via photos for respondents. 

\subsection{Impacts of Taking Medical Selfies}
By inspecting their bodies as part of the study procedure, participants raised their awareness of their bodily health status. Four participants noticed that their skin was dry, mostly on the feet. For example, one participant discovered new moles while checking on another mole that they had not checked in years (P4). Other features participants noted were stretch marks, sun burns and peeling callus. The process of self-examination made participants check on body parts they would not normally look at in their daily life.

Individuals were reminded of past events associated with certain body lesions: \say{I had surgery a few years ago. [...] I had forgotten I had this scar. And as I now saw it again, I was reminded of this surgery. A bit... It was not bad the feeling, but the surgery was present again. Not so pleasant} (P2). in some cases strong feelings in relation to past events were evoked during the capture process:
\say{I also see a spot there. I don't know if it's from a rash I had long time ago. [...] Indeed, something of it remained and then I remembered everything. Yes exactly. That was very tough for me back then, because I had a great birthday party, spent a lot of money and in South Africa, where I come from, [...] and then I went to the doctor [...] and spent a lot of money} (P4).

During the process of taking the photos, four participants (33\%) mentioned the aesthetic dimensions of the photos. Two participants (17\%) stated they should get a pedicure done after examining their feet. Armpits caused feelings of disgust and shame, as expressed by two individuals (17\%): \say{Armpits disgust me. How can it be this hard to talk about it? It somehow annoys me, and I do not think armpits are pretty, on anybody} (P3). A second participant stating, \say{so I probably have to photograph there, well now it just comes to my mind that I should have shaved my armpits before this} (P8). One participant reflected on the fact that the aesthetic character of medical photos significantly differs from sending photos to other people: \say{Yes, it's stupid, I know, because a dermatologist looks at it in a different way, but somehow you always want to look as good as possible} (P4). Another participant stated that they wanted to take the photos from an angle and in the lighting where their body parts look the best (P10).

\subsection{Data Privacy Concerns}
Data privacy was an important topic for most participants (Table \ref{tab:themes}). A major concern, mentioned by a third of participants (33\%) was that photos taken with the smartphone would automatically be uploaded to the internet via cloud services embedded in the devices. Three individuals (25\%) were concerned that, as they appeared in the photo gallery, they might accidentally show the pictures to other people when showing other images. To prevent possible tracking, one participant (P11) reported that they typically remove all the metadata from photos before sharing them. One participant related a prior situation, not in the context of the study, where they sent a sensitive photo, to the wrong person: \say{I was actually taking a picture for my mom when I had a strange mark on my breast and I accidentally sent it to my aunt's husband and damn, I felt ashamed. Sure, I removed it right away, but since then I have been thinking every day that, damn it, if he saw it there, it would be quite embarrassing} (P3).

\section{Discussion}
In this section we discuss our findings in relation to our research questions and prior works, present design guidelines and describe limitations of our work.

Our RQ1 is related to the emotional impacts associated with the process of taking medical selfies and how they differ between body parts. Not surprisingly, emotional impacts of medical selfies became apparent especially when photographing intimate areas of the body. Both the survey respondents and study participants similarly expressed discomfort when considering, or actually capturing, images of their buttocks and genitals. Gender differences were identified in the range of body parts that respondents had captured and sent as medical selfies, with women photographing a broader set of parts, but not in the overall frequency of sending such photos. While only a small subset of our sample had prior experience with sharing medical images with a doctor, sharing medical selfies with close friends and family is a common practice, although typically done infrequently.

Taking medical photographs of different parts of our bodies without using any filters or aiming for an aesthetic photograph, differs greatly from the current mainstream selfie-culture ~\cite{lonergan2019me,cohen2018selfie,sung2016we}, potentially causing affective responses that should be considered in the development of imaging applications for medical health purposes. However, interestingly, for some participants the aesthetic dimensions of selfies were also present in the medical context of our study. Participants were concerned about their appearance in medical selfies, even though they were aware of the sole medical purpose. Similar to ~\cite{mclean2015photoshopping} who report on staging behavior, e.g. setting up the environment to take selfies, participants in our study put efforts into making the photos look aesthetic. We derive two implications from this finding. First, this behavior could impact image quality, e.g. when participants prioritize aesthetics over accurate representation of skin lesions. However, secondly, as posting selfies encourages patients to take ownership of their condition \cite{tembeck2016, burns2019}, the creation of aesthetic medical selfies should not be discouraged. Hence, designers and developers of mHealth apps should not reduce medical selfies to a single functional purpose of disease diagnosis, but also accept them as a means of patients' self expression and enable users to take aesthetically pleasing medical images.

Data privacy concerns were one of the most important reasons for participants not to take photos of their genitals. To build trust in mHealth applications, data privacy should be considered a high priority. Besides the actual implementation of standards such as end-to-end encryption, these measures should be clearly communicated to the users, e.g. by continuously displaying a disconnection state from cloud storage. As proposed by the trustworthiness model by Akter et al.~\cite{akter2011trustworthiness} and consistent with Lupton and Jutel~\cite{lupton2015s}, disclaimers in the context of diagnoses increase user trust. To prevent accidental leakage of sensitive images, medical selfie images should be stored separately from other images and require additional confirmation when sharing them.

Individuals in the survey and, to a limited extent, in the study had a preference for familiar doctors to diagnose their medical selfies. Therefore, we suggest that mHealth solutions support the patient-doctor relationship by displaying information about the diagnosing doctor, e.g. through photos and a description text. This design idea supports the concept of alternating between telemedicine and personal patient-to-doctor interactions proposed by Scott et al. ~\cite{scott2018evaluating}.

Our second research question, RQ2, explores the physical and practical challenges people face when taking medical selfies and how mHealth apps can be designed to overcome them. Participants had difficulties in taking high quality medical selfies due to uncertainty in device and body positioning, poor focus and a lack of good lighting, factors which have also been observed in prior work ~\cite{ashique2015clinical}.
Providing assistance in the form of model images for the user to aim to replicate or visual/auditory guidance before and during the capture process may provide solutions to mitigate some of these challenges~\cite{Diethei:2018}. Solutions such as the automated shutter release of the Miiskin app\footnote{https://miiskin.com/app/} may be effective in ensuring image quality, but without appropriate guidance may become frustrating for the user.

In terms of physical demand, the body parts on the back side of the body, the buttocks and shoulder blades, were the hardest to capture. While it could be argued that for those body parts people can ask someone else to take the photos, this is not always possible and especially difficult for the buttocks, which are among the most sensitive body parts. Potential solutions are using two mirrors~\cite{friedman1991melanoma} or selfie-sticks~\cite{Arif2017}. Furthermore, the app should enable images to be taken without the need to manually press a shutter button.

Our findings provide insights to support application developers and HCI practitioners to take into account the role of emotional and physical experiences when designing digital health technologies for skin self-examination (Table \ref{tab:guidelines}). As a result we hope this will lead to the development of digital health technologies that encourage, rather than inhibit, people to use them, and to increase their effectiveness in disease diagnosis and treatment.

\begin{table*}[b]
\centering
\begin{tabular}{ l | m{9cm} }
\textbf{Issue}  & \textbf{Design Guideline} \\
\hline
Device positioning & Provide model images for the user to aim to replicate.  \\
Device positioning & Provide a "mirror mode" to guide participants to use a mirror to take photos of the back side of their body. \\
Image acquisition & Enable images to be taken without the need to manually press a shutter button.\\
Appearance in Medical Selfies & Enable users to take aesthetically pleasing medical images. \\
Data Privacy Concerns & Continuously display disconnection state from cloud storage. \\
Data Privacy Concerns & Store the medical selfie images separately from other images and require additional confirmation when sharing them. \\

\end{tabular}
\vspace{3pt}
\caption{Guidelines for the design of medical selfie mHealth apps derived from observations in the in-depth user study.}~\label{tab:guidelines}
\end{table*}

\subsection{Methodological Notes}
In general, user study procedures such as ours, which require participants to independently follow multiple instructional steps in their own homes, face challenges of data quality and completeness. Here we discuss the challenges in our study, with the aim of providing guidance for future researchers to minimise such issues.
Firstly, there was one occurrence where a participant almost accidentally sent the study photos to the researchers (instead of only the audio and video recordings). Had this happened, it would have created privacy and ethical issues that were not covered by the ethical plan for the study, and that the researchers were unprepared for. Planning how to minimise the potential for such human errors, and having a recovery plan in place would be beneficial in future studies of a similar nature. Second, three participants dropped out of the study due to problems with the study procedure; two individuals reported that they felt overwhelmed, while one did not manage to send the recordings, likely due to usability issues with using their smartphone to record audio. Our procedure could have been improved by providing an integrated approach, i.e. a study-specific application. This would also increase data privacy of the photos since cloud upload and photo gallery `leaks' can be prevented by default. Additionally, the recordings could be sent to the researchers in a safe, i.e. encrypted, way instead of through e-mails. 

Overall, we regard our study protocol as suitable for the purpose of collecting data about medical selfies. Especially the audio recorded thinking aloud method led to a lot of valuable in-situ insights which would not have been captured with a retrospective interview or questionnaire. However, the think aloud method led to some participants feeling that researchers were somehow present during the photo taking and not completely immersing in the imaginary scenario of taking photos for medical purposes, e.g. commenting, \say{[I] could not completely forget that it's not for my doctor, but for a study} (P1). As the data collection process was already in a rather uncontrolled setting, a less unobtrusive experimental approach would have been difficult to achieve.

The data collected in the survey and the study originated from three distinct populations. While the survey respondents were located in the USA, study participants resided in Finland and Germany. The cultural differences between these groups were not analyzed. We wanted to achieve a wide spread of study sites across the western worlds. We had to weigh up in which geographical areas we could conduct the study. Instead of studying only one western country, we chose Finland, Germany and the USA, as they have different health care systems but all share the western culture. Since the authors had access to users in Finland and Germany, we decided to conduct the study in these countries. We noticed differences in the healthcare systems that impacted the user study in particular. While in Germany most patients have a nominated `house doctor', in Finland the relation between a patient and an individual doctor is less close. Therefore, sending photos to a specific and familiar doctor was regarded more important for the German participants. We also noted that Finnish participants had more experience with telemedicine. Our combination of rich in-depth interviews plus a larger sample survey provides both depth and breadth to our findings, which we believe supports the generalisation of our findings. 


\subsection{Limitations and Future Work}
The international nature of the study, conducted in Finland and Germany, made it necessary to translate most of the collected data from the participants' native languages to English as a working language because not all the study's researchers spoke both German and Finnish. According to \cite{lamnek2005qualitative}, every translation is an interpretation and as such is neither objective nor neutral. Hence, the translation should be considered as part of the research process. Since we conducted the analysis on translated versions of the audio, video and interview data, some of the affective clues may have become distorted from their meaning in their original language.

In future, we plan to conduct further studies on the topic, for example investigating the role of discussion forums, such as Reddit, in the context of SSE.

\section{Conclusion}
We collected two datasets, a survey (n=100) and an in-depth user study (n=12), exposing user perceptions of taking selfie photographs of skin conditions for diagnosis. 
When taking medical selfies of intimate body regions, people felt emotional discomfort and expressed data privacy concerns. For some participants, the aesthetics of the photos played an important role, even though they recognized their medical purpose. Participants faced practical problems with positioning to photograph their back and buttocks. Issues of lighting and camera focus were noted as challenges by participants. Our findings highlight the importance of addressing practical and usability issues as well as managing privacy in the user interface of e-health apps, building trust between users and the system. Methodologically, our approach of data collection, with participants recording their thinking aloud audio while capturing the medical selfies, and later video recordings provided a good trade-off between maintaining privacy and data quality.


\section{Acknowledgements}
We would like to thank the Volkswagen Foundation for funding this research. We gratefully acknowledge the support of the Leibniz ScienceCampus Bremen Digital Public Health (lsc-diph.de), which is jointly funded by the Leibniz Association (W4/2018), the Federal State of Bremen and the Leibniz Institute for Prevention Research and Epidemiology – BIPS. Lastly, this work was supported through KI-SIGS (grant 01MK20012).

\bibliographystyle{ACM-Reference-Format}
\bibliography{refs}

\end{document}